\documentclass[superscriptaddress,reprint,amsmath,amssymb,aps,prl]{revtex4-1}
\usepackage{graphicx}
\usepackage{bm}
\usepackage{color}
\usepackage{hyperref}
\hypersetup{colorlinks,citecolor=blue}
\newcommand{\ket}[1]{\mbox{$|#1\rangle$}}

\newcommand{\+}{$^+$}

\begin{document}

\title{Cooling of a Zero-Nuclear-Spin Molecular Ion to a Selected Rotational State}

\author{Patrick R. Stollenwerk}
\thanks{Equally contributing authors}
\affiliation{Argonne National Laboratory, Lemont, Illinois 60439, USA}
\author{Ivan O. Antonov}
\thanks{Equally contributing authors}
\affiliation{Department of Physics and Astronomy, Northwestern University, Evanston, Illinois 60208, USA}
\author{Sruthi Venkataramanababu}
\affiliation{Graduate Program in Applied Physics, Northwestern University, Evanston, Illinois 60208, USA}
\author{Yen-Wei Lin}
\affiliation{Department of Physics and Astronomy, Northwestern University, Evanston, Illinois 60208, USA}
\author{Brian C. Odom}
\email{b-odom@northwestern.edu}
\affiliation{Department of Physics and Astronomy, Northwestern University, Evanston, Illinois 60208, USA}

\date{\today}

\begin{abstract}
We demonstrate rotational cooling of the silicon monoxide cation via optical pumping by a spectrally filtered broadband laser. Compared with diatomic hydrides, SiO\+ is more challenging to cool because of its smaller rotational interval. However, the rotational level spacing and large dipole moment of SiO\+ allows direct manipulation by microwaves, and the absence of hyperfine structure in its dominant isotopologue greatly reduces demands for pure quantum state preparation.  These features make $^{28}$Si$^{16}$O\+ a good candidate for future applications such as quantum information processing. Cooling to the ground rotational state is achieved on a 100 ms time scale and attains a population of 94(3)\%, with an equivalent temperature $T=0.53(6)$ K. We also describe a novel spectral-filtering approach to cool into arbitrary rotational states and use it to demonstrate a narrow rotational population distribution ($N\pm1$) around a selected state.

\end{abstract}
\maketitle

\section{Introduction}

A broad range of physics and chemistry interest motivates developing tools for robust control over molecular internal degrees of freedom. Applications include study of cold collisions~\cite{zhang2017cold,dorfler2019long}, quantum state-dependent chemistry~\cite{bell2009ultracold,hutson2010ultracold,balakrishnan2016perspective}, astrochemistry~\cite{wm2008low}, many-body physics~\cite{yan2013observation,hazzard2014many,schmidt2015rotation,schmidt2016deformation}, quantum information processing ~\cite{ni2018,hudson2018dipolar,campbell2019dipole}, and precision spectroscopy~\cite{safronova2018search, Cairncross2017, Andreev2018, Alighanbari2020, Chou2020}.
Molecules, in contrast to free atoms, have rotational and vibrational degrees of freedom via their chemical bonds. On one hand, these extra degrees of freedom extend the scope of possible control and provide the rich structure that generates their appeal. On the other hand, their level structure can be quite complicated and thus challenging for state control. Despite this increased complexity, substantial progress on state preparation of molecules has been made in the past decade using several techniques including optical pumping~\cite{viteau2008optical,staanum2010rotational, schneider2010all, cournol2018rovibrational, lien2014broadband}, buffer-gas cooling~\cite{Rellergert2013,hansen2014efficient}, state-selective photoionization~\cite{tong2010sympathetic}, projective preparation using quantum logic~\cite{chou2017preparation}, supersonic expansion of molecular beams~\cite{shagam2013sub}, and photoassociation~\cite{ni2008high}.

Trapped molecular ions in Coulomb crystals can be isolated from the environment and are well-suited for precision spectroscopy, quantum information processing, and other applications requiring uninterrupted dynamics over long time scales. State preparation by optical pumping allows rapid resetting of the molecular state, often desired in these applications.  Optical pumping of trapped molecular ion rotations has previously been demonstrated for diatomic hydrides~\cite{staanum2010rotational,schneider2010all,lien2014broadband}. However, the non-zero nuclear spin of hydrogen (or deuterium) couples with the rotational degree of freedom and any nuclear spin of the other atom, making optical pumping to a pure state still a challenge.  The only demonstration of simultaneous rotational and nuclear spin optical pumping achieved $\sim$20\% hyperfine state purity of HD$^+$ in a few tens of seconds~\cite{bressel2012manipulation}.  A somewhat different quantum-logic approach determines with high confidence into which pure state a single trapped CaH\+ ion is non-destructively projected, but the \emph{a priori} probability of being in a given rotational manifold is currently limited to the $\sim$10\% thermal population~\cite{chou2017preparation}.

In contrast to hydrides, high natural-abundance oxide isotopologues exist where both atoms have nuclear spin $I=0$, circumventing the challenge of hyperfine structure in quantum state preparation.  Taking advantage of this simplification, we demonstrate here broadband optical pumping of $^{28}$Si$^{16}$O\+ to its ground rotational level, with well-defined total angular momentum, on a timescale of 100 ms with 94(3)\% fidelity.  We also demonstrate, although with lower state purity, optical pumping to a selected rotational state with $N>0$.

The rotational spacing of oxides (43 GHz for the lowest SiO\+ interval) is smaller than the few to several hundred GHz typical of hydrides.  This makes cooling them more challenging, both because the optical pumping spectrum is more congested and because more levels are thermally populated; 95\% of the 300 K population is distributed over the rotational states $N\le30$ for SiO\+, compared with $\sim$10 for typical hydrides.  Although technically more challenging for cooling, the smaller rotational interval of oxides is favorable for future applications.  SiO\+ has a sizeable body-frame dipole moment of $>4$ Debye~\cite{werner_ab_1982,cai1999ab, chattopadhyaya2003electronic, li2019laser}, so coherent rotational transitions can be driven with convenient microwave sources.  Also rotational transitions of oxides are further from the peak of the 300 K blackbody spectrum, so ultimate lifetimes and coherence times for polar oxides can be one to three orders of magnitude longer, depending on other limiting factors.

\begin{figure*}
\includegraphics[width=\textwidth]{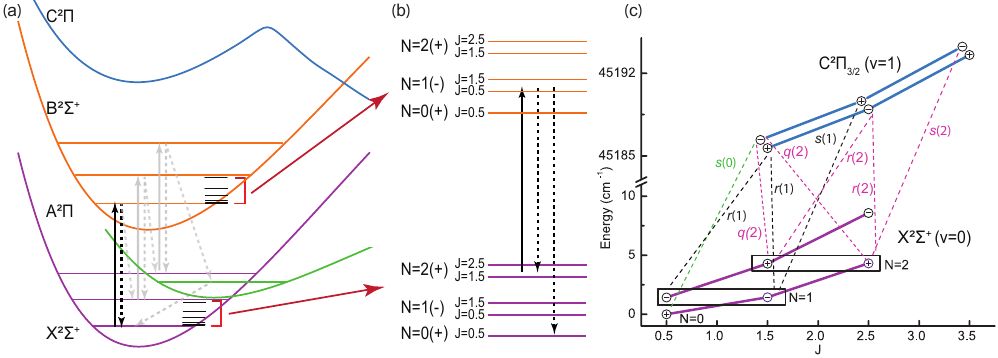}
\caption{\label{fig:levels} Transitions for pumping and state readout.  \textbf{(a)} Potential energy curves (not to scale) for the lowest three electronic states along with the higher-lying C state, with representative vibrational and rotational substructure. Black arrows show the primary pumping excitation and spontaneous emission channel, and gray arrows a possible parity flip sequence.  \textbf{(b)} Rotational pumping from $N=2 \rightarrow 0$.  \textbf{(c)} Dissociative transitions from $N={0,1,2}$. }
\end{figure*}

\section{Broadband Optical Pumping}
At 300 K, 99.6\% of SiO\+ molecules are in the ground vibrational state $v=0$, but population is spread over $\sim$30 rotational levels.  We desire rotational pumping which minimizes unwanted incidental vibrational excitations.  Several groups have noted the advantage of optical pumping with the class of molecules whose ground and excited electronic states have similar bond equilibrium distances, i.e. molecules with nearly diagonal Franck-Condon Factors (FCFs)~\cite{DiRosa2004,sofikitis2009molecular,nguyen2011prospects,nguyen2011challenges,cournol2018rovibrational}. Such molecules have electronic excitation largely decoupled from vibrational excitation. In SiO\+ the diagonal FCFs of the $B$-$X$ transition (Fig~\ref{fig:levels}(a)) allow on average more than 30 optical pumping cycles before $v$ changes~\cite{stollenwerk2017electronic}.

Diagonal FCFs also imply the states have nearly identical rotational constants, resulting in a spectrum well separated according to angular momentum selection rules. A broadband laser with relatively simple spectral filtering tuned to a diagonal transition in AlH\+ was used to achieve high-fidelity cooling to the ground rotational state~\cite{lien2014broadband}. Since the rotational constant of SiO\+ is an order of magnitude smaller, rotational cooling of this new species to a similar degree requires significantly better spectral filtering.

\begin{figure}
\includegraphics[width=3.375in]{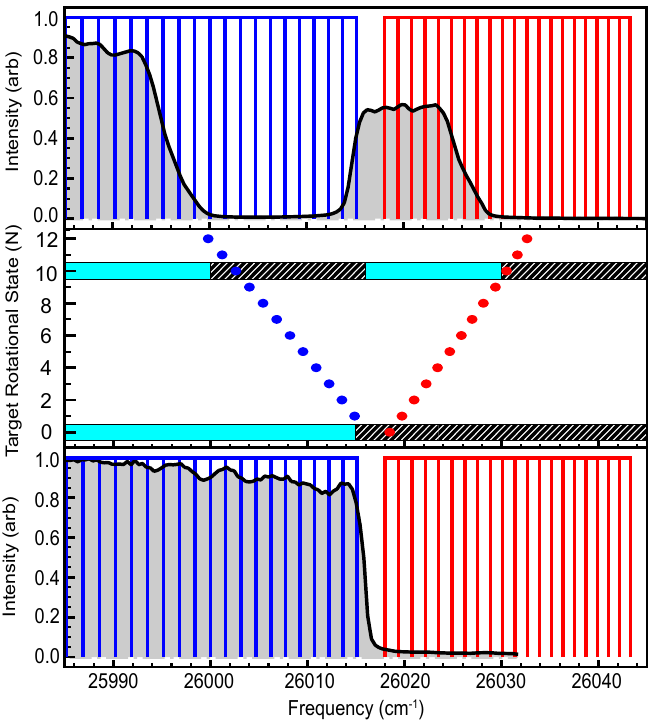}
\caption{\label{fig:pshape} \textbf{(Center)} Spectral masks overlayed onto the Fortrat diagram of the SiO\+ $B - X (0,0)$ transition. $P$ and $R$-branch transitions are indicated by blue and red dots respectively. The blocked spectrum for each mask is indicated by the dark hashes. The measured optical pumping spectra which have been filtered to target the ground rotational state \textbf{(Bottom)} and $N=10$ \textbf{(Top)} are shown with the relevant $P$ and $R$-branch transitions indicated by blue and red vertical lines. }
\end{figure}

State preparation of SiO\+ was achieved by spectrally filtering a frequency doubled Spectra-Physics MaiTai HP laser tuned to the $B^2\Sigma^+ - X^2\Sigma^+$ electronic transition near 385 nm. Spectral filtering was done by using the 2-$f$ configuration of the pulse-shaping setup described in~\cite{lin2016high}. For ground state preparation the spectral filtering mask requires pumping of only the $P$-branch transitions ($\Delta N = -1$) which is accomplished by blocking the high frequency components at the Fourier plane with a razor blade (Fig.\,\ref{fig:pshape})~\cite{lien2011optical,lien2014broadband}.

To extend preparation to arbitrary $N>0$ rotational states, a mask on the $P$-branch must be introduced to only pump down to the target state, and the mask on the $R$-branch needs to be shifted to allow pumping up to the target state (Fig.\,\ref{fig:pshape}).  This is accomplished with the removal of a band in the middle of the spectrum in addition to the removal of the high frequency components. This band is filtered using a thin metal ribbon (0.038~$\times$~3~mm) whose profile is adjusted by rotating to match the required bandwidth at the Fourier plane. In this way, each rotational level is exclusively pumped toward the target state, which is intentionally left dark.

\section{Trapping and Detection}
Quantum state control experiments were performed at room temperature under ultra-high vacuum conditions ($7(4)\times 10^{-10}$ Torr). For each data point a sample of 10 to 100 SiO\+ was co-loaded with 500 to 1000 laser cooled barium ions in a linear Paul trap. Loading of SiO\+ was performed using the SiO $A^1\Pi - X^1\Sigma^+(5,0)$ 1+1 REMPI transition following ablation of a solid SiO sample~\cite{stollenwerk2017ip}. Translational energy is rapidly cooled sympathetically by Ba\+, however molecular internal degrees of freedom are decoupled from translational motion. We expect to load SiO\+ between $N=4$ and $N=15$~\cite{stollenwerk2017ip}.  The interval between loading and dissociation is typically 30 s at which time, without optical pumping, BBR and spontaneous emission have redistributed population from a single $N$ into nearby levels, but full thermal equilibration has not yet occurred.  We do not observe significant population in $N=0$ without optical pumping.

The trapped SiO\+ were detected using an in-situ laser cooled fluorescence mass spectrometry (LCFMS)~\cite{baba2001laser} technique. Briefly, the radial secular motion of the SiO\+ was resonantly excited using a low amplitude (0.5-1 V) RF waveform applied to one of the radial trapping rods. Motional excitation causes Doppler broadening of the Ba\+ resonance and a decrease of Ba\+ fluorescence in proportion to the number of SiO\+ in the trap.

State detection of SiO\+ was performed destructively using single-photon resonance-enhanced photodissociation spectroscopy via the predissociative $C^2\Pi$ state, which our preliminary linewidth measurements indicate has a lifetime of order of several hundred picoseconds for $v=0$ and less for higher $v$.  These lifetimes are sufficiently long to provide rotational resolution and sufficiently short for efficient dissociation. The state was previously reported in theoretical studies as the $2^2\Pi$ state~\cite{cai1999ab, chattopadhyaya2003electronic, honjou2003ab, shi2012mrci,li2019laser}. It had not been observed experimentally prior to this work. A manuscript detailing spectroscopy of the $C^2\Pi$ state is currently in preparation~\cite{Science2020}.

A pulsed dye laser with frequency-doubled output near 226 nm was used for dissociation. Predissociation of SiO\+ via the $C^2\Pi$ state leads to Si\+ + O products, and we monitored the LCFMS SiO\+ signal to measure the dissociated fraction. We performed two types of measurements to characterize state control.

The first method uses slow steady-state depletion, where the rate constant yields relative populations. The LCFMS signal was monitored while optical pumping and concurrently firing the 10 Hz dissociation laser tuned to frequency $f$ for 30 s. If dissociation is slow enough, steady-state population is maintained, and the number of SiO\+ molecules as a function of pulse number $m$ is given by $N_{SiO^+} = N_0 e^{-\Gamma(f) m}$.  Each pulse dissociates a fraction of remaining molecules given by
\begin{equation}
\label{ratepop}
d(f) = p_N(f)\, n_N \approx -\frac{d N_{SiO^+}}{d m} \frac{1}{N_{SiO^+}} = \Gamma(f),
\end{equation}
where $p_N(f)$ is the probability per pulse of dissociating a molecule which is in the probed state $N$, and $n_N$ is the fraction of remaining molecules in that state. To ensure that Eqn.\,\eqref{ratepop} is valid, we require $p_N(f) \ll 1$ and $\Gamma \ll \Gamma_{eq}$, where $\Gamma_{eq}$ is the equilibration rate of the probed state.  Experimentally, we reduce the dissociation laser fluence until these conditions are met. A fit to the LCFMS decay at each $f$ yields the dissociation spectrum $d(f)$, in which the peak heights depend both on line strength and population. Although these spectra show only relative populations, this first method provides a good signal-to-noise ratio (SNR), since the entire sample contributes toward statistics even for probed states with low population. Also, spectra can be taken with constant SNR over a dynamic range of more than two orders of magnitude.

\begin{figure}
\includegraphics[width=3.375in]{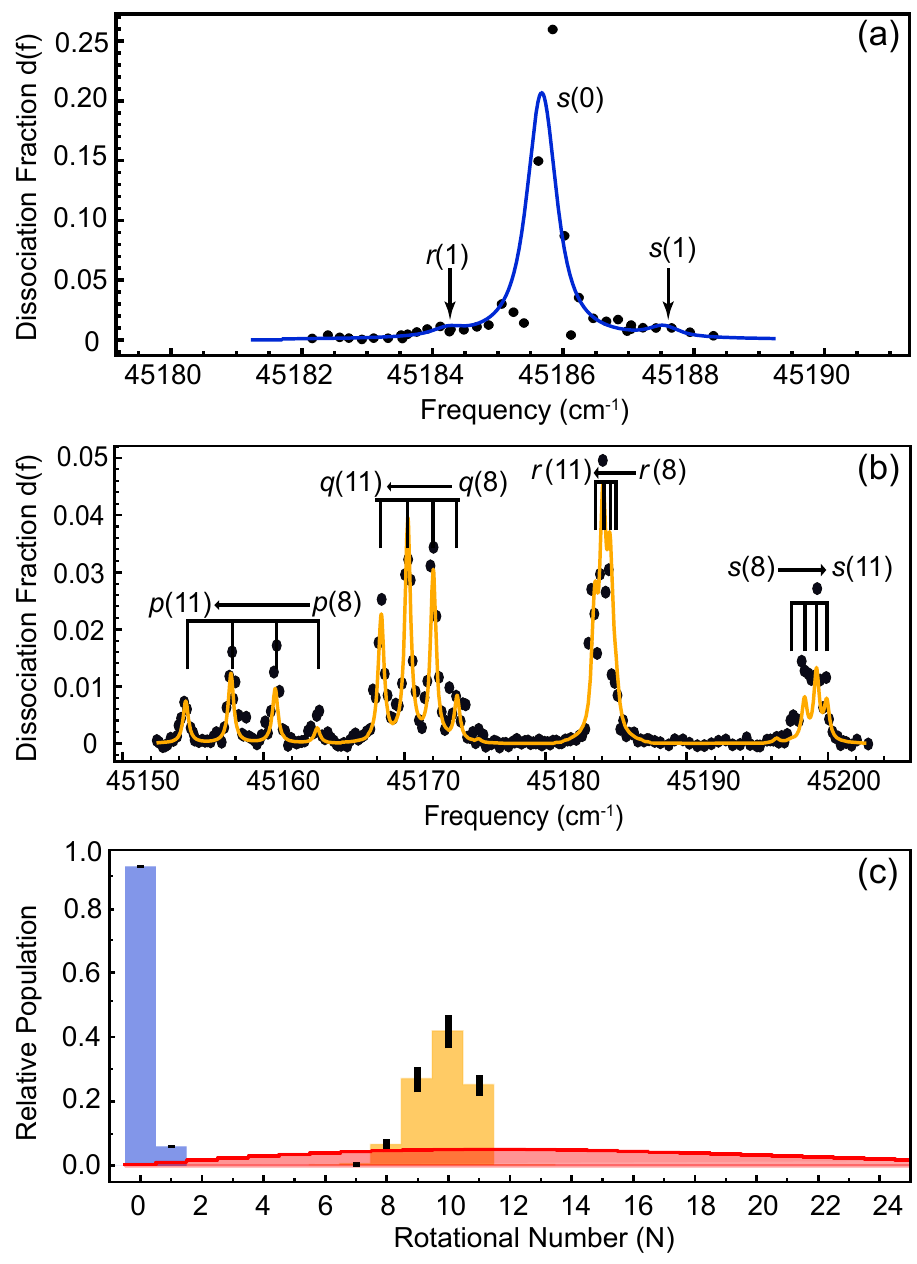}
\caption{\label{fig:N10pop}  Dissociation spectra of SiO\+ optically pumped \textbf{(a)} towards $N=0$ and \textbf{(b)} $N=10$. \textbf{(c)} Relative rotational state populations inferred from the spectra and their $\pm 1 \sigma$ uncertainties, along with a 300 K thermal distribution (red). }
\end{figure}

The second method is a single-shot depletion technique ~\cite{staanum2010rotational,schneider2010all,lien2011optical}, which yields absolute populations. We recorded the LCFMS signal before and after a single intense pulse tuned to dissociate from state $N$, where $p_N(f)\sim 1$. Because the predissociation lifetime of the upper state is much shorter than the 10 ns pulse duration, 100\% dissociation probability is achievable. The fractional population in $N$ is given by $F_N = (D_i - D_f)/D_i$, where $D_i$ and $D_f$ are the (positive-valued) LCFMS fractional fluorescence dips before and after the dissociation pulse.

\section{Relative Cooling Efficiency}
Fig.~\ref{fig:levels}(c) shows the dipole-allowed $\ket{X^2\Sigma^+,v=0} \rightarrow \ket{C^2\Pi_{3/2},v=1}$ dissociative readout transitions. This vibronic transition was chosen because it exhibits good separation between lines originating from $N=0$ and $N=1$ as well as from other isotopologues.  Each originating $N$ has up to four resolvable lines labeled as $x(N)$.  The branch type $x$ is characterized by $\Delta N = (J'+1/2)-N$, where $J'$ is the upper rotational quantum number; e.g. $^sR_{21}(0.5)$ is denoted $s(0)$.

Fig.\,\ref{fig:N10pop} shows the spectrum after the population has been pumped toward $N=0$. We simulated the spectrum using PGOPHER~\cite{western2017pgopher} and fit the spectral envelope to obtain a ratio of population in $N = 1$ to $N = 0$ of 0.075(3). We also demonstrate cooling into an excited rotational state by applying the spectral mask for $N=10$ (Fig.\,\ref{fig:pshape}). Both the $N = 0$ and $N = 10$ spectra are in sharp contrast with a thermal distribution at 300 K.

Although the steady-state analysis technique does not directly yield
absolute populations, some qualitative conclusions can be drawn about populations in other states.  Scans searching for transitions originating from $N\ge2$ of $v=0$ and for any $N$ of $v=1$ did not show any discernable peaks.  The states in the $A^2\Pi$ manifold are too short-lived for significant population accumulation.  A quantitative measurement setting bounds on these and other populations is discussed in the section below.

\section{Absolute Cooling Efficiency and Timescale}
\begin{figure}
\includegraphics[width=3.375in]{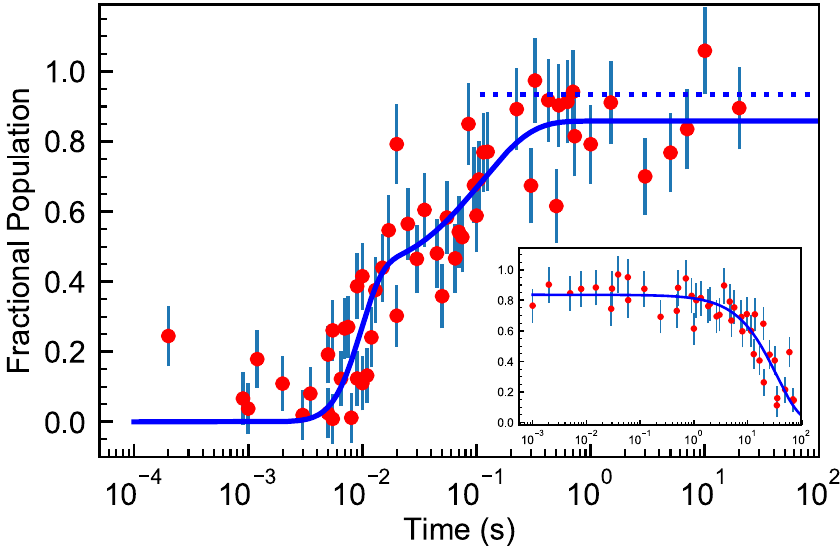}
\caption{\label{fig:cool} Single-shot analysis of $N=0$ population versus optical pumping time.  Error bars account for technical noise and SiO\+ counting statistics.  Since LCFMS involves difference measurements, points and error bars outside of the range of 0 to 1 are expected.  The solid line is a fit to the model, and the horizontal line indicates the corrected asymptote accounting for systematic shifts. \textbf{(Inset)}
Measured longevity of molecules prepared in $N=0$, after the pumping laser is turned off.}
\end{figure}

Fig.\,\ref{fig:cool} shows the measured population accumulation in $N=0$ when pumping toward that state, analyzed using the single-shot method. Here, we used the $\ket{X^2\Sigma^+,v=0,N=0} \rightarrow \ket{C^2\Pi_{1/2},v=0,J=0.5}$ transition at 44044.5 cm$^{-1}$. Technical noise, which dominates over SiO\+ counting noise, is primarily due to laser fluctuations affecting Ba\+ cooling efficiency and fluorescence. The anomalous point at very short times is understood to be a spurious signal from population in $N=11$, which has an overlapping line.

Two time scales are present. The faster time scale is for cooling of the separate parity populations independently. Photon absorption and then spontaneous emission on the diagonal $B$-$X$ transitions is a parity-preserving process, so approximately half of the population is unable to be directly pumped into the even-parity ground rotational state.  The slower time scale is determined by the rate of parity flips. The details of this process have not been determined, but one possible pathway for obtaining the requisite odd number of electric dipole transitions is shown in Fig.~\ref{fig:levels}(a) (also see Supplemental Material).

An analytic model (see Supplemental Material) is used to fit the data.  To assess absolute cooling efficiency, an offset must be applied to correct systematic shifts from reaction with the background hydrogen, isotopic impurities, and partial overlap of the laser linewidth with other dissociating transitions. We conclude that the steady-state $N=0$ absolute population fraction of $^{28}$Si$^{16}$O\+ is 0.94(3), which corresponds to a temperature of 0.53(6) K.  This result is in good agreement with the relative population analysis.

Thermalization of SiO\+ out of $N=0$, after the cessation of optical pumping, is shown in the inset of Fig.\,\ref{fig:cool}. The population loss is well fit by an exponential decay with a time constant of 35(4) s. This time scale is much faster than blackbody-induced pure rotational (400 s) or vibrational excitations (380 s), and also faster than the observed reaction rate with H$_2$ ($\sim$600 s).  However, it could be consistent with inelastic collisions with H$_2$, which have a Langevin collision time of 40(20) s, with uncertainty coming from the H$_2$ pressure. Blackbody redistribution via the $A$ state is also a possible mechanism, expected to occur on a time scale of 70 - 130 s given the predicted $A$ state lifetimes~\cite{nguyen2011prospects,li2019laser}.

\section{Conclusions}
This work demonstrates the extension of a broadband rotational cooling technique from a trapped diatomic hydride to an oxide. Furthermore, we show that the class of diagonal molecules amenable to rotational cooling can be extended to include those with an intermediate electronic state not involved in the dominant optical pumping cycle.

Higher resolution spectral filtering is possible, for example by using a virtual imaged phased array (VIPA) where sub-GHz resolution has been achieved~\cite{willits2012line}. VIPA could significantly enhance the preparation fidelity of the $N=0$ and $N=10$ states shown here. Consequently, it might be advantageous to use $N>0$ states for quantum information processing, since dominant decoherence mechanisms are reduced for higher rotational states~\cite{hudson2018dipolar}. Optical pumping to $N\gg0$, not explored in this work, can also be useful for spectroscopic studies and can provide insights into the molecular Hamiltonian at high energies~\cite{Science2020}.

The limiting timescale is currently the parity cooling step.  Electronic decay from $A$ to $X$ is predicted to be of order 5 ms~\cite{li2019laser}, thus we expect that the cooling rate could be increased by more than an order of magnitude with increased spectral fluence of the pump laser.  An alternative would be to use microwaves to drive parity flips, for instance at 86~GHz connecting $N=1$ with $N=2$~\cite{scholl_laser-rf_1995} for cooling to $N=0$.  Use of a microwave drive could equalize the timescales for cooling the two parities, and with a more intense cooling laser $^{28}$Si$^{16}$O\+ population could be cooled to a pure internal quantum state in $<10~\mu$s, limited by the spontaneous emission rate of $B$. Further refinements may enable fluorescence imaging or direct Doppler cooling of SiO$^+$~\cite{nguyen2011prospects,li2019laser} and could help realize multi-ion molecular clocks with canceling Stark and second order Doppler shifts~\cite{stollenwerk2018optical}.

We have demonstrated straightforward pumping of $^{28}$Si$^{16}$O\+ to a state of well-defined angular momentum.  Since $^{28}$Si$^{16}$O\+ also has microwave-accessible rotational transitions for quantum manipulation, this species could play a similar role in a wide range of applications, as do only a relatively small handful of popular atomic species.

\begin{acknowledgments}
We gratefully acknowledge Jason Nguyen for proposing rotational cooling of SiO\+, along with his and David Tabor's early work on SiO\+ in the lab. Development of dissociative analysis techniques were funded by NSF Grant No. PHY-1806861, LCFMS techniques were funded by ARO Grant No. W911NF-14-0378, and optical state control techniques were funded by AFOSR Grant No. FA9550-17-1-0352.
\end{acknowledgments}


\begin{thebibliography}{50}
\expandafter\ifx\csname natexlab\endcsname\relax\def\natexlab#1{#1}\fi
\expandafter\ifx\csname bibnamefont\endcsname\relax
  \def\bibnamefont#1{#1}\fi
\expandafter\ifx\csname bibfnamefont\endcsname\relax
  \def\bibfnamefont#1{#1}\fi
\expandafter\ifx\csname citenamefont\endcsname\relax
  \def\citenamefont#1{#1}\fi
\expandafter\ifx\csname url\endcsname\relax
  \def\url#1{\texttt{#1}}\fi
\expandafter\ifx\csname urlprefix\endcsname\relax\def\urlprefix{URL }\fi
\providecommand{\bibinfo}[2]{#2}
\providecommand{\eprint}[2][]{\url{#2}}

\bibitem[{\citenamefont{Zhang and Willitsch}(2017)}]{zhang2017cold}
\bibinfo{author}{\bibfnamefont{D.}~\bibnamefont{Zhang}} \bibnamefont{and}
  \bibinfo{author}{\bibfnamefont{S.}~\bibnamefont{Willitsch}}, in
  \emph{\bibinfo{booktitle}{Cold Chemistry}} (\bibinfo{year}{2017}), pp.
  \bibinfo{pages}{496--536}.

\bibitem[{\citenamefont{D{\"o}rfler et~al.}(2019)\citenamefont{D{\"o}rfler,
  Eberle, Koner, Tomza, Meuwly, and Willitsch}}]{dorfler2019long}
\bibinfo{author}{\bibfnamefont{A.~D.} \bibnamefont{D{\"o}rfler}},
  \bibinfo{author}{\bibfnamefont{P.}~\bibnamefont{Eberle}},
  \bibinfo{author}{\bibfnamefont{D.}~\bibnamefont{Koner}},
  \bibinfo{author}{\bibfnamefont{M.}~\bibnamefont{Tomza}},
  \bibinfo{author}{\bibfnamefont{M.}~\bibnamefont{Meuwly}}, \bibnamefont{and}
  \bibinfo{author}{\bibfnamefont{S.}~\bibnamefont{Willitsch}},
  \bibinfo{journal}{Nat. Commun.} \textbf{\bibinfo{volume}{10}},
  \bibinfo{pages}{1} (\bibinfo{year}{2019}).

\bibitem[{\citenamefont{Bell and P.~Softley}(2009)}]{bell2009ultracold}
\bibinfo{author}{\bibfnamefont{M.~T.} \bibnamefont{Bell}} \bibnamefont{and}
  \bibinfo{author}{\bibfnamefont{T.}~\bibnamefont{P.~Softley}},
  \bibinfo{journal}{Mol. Phys.} \textbf{\bibinfo{volume}{107}},
  \bibinfo{pages}{99} (\bibinfo{year}{2009}).

\bibitem[{\citenamefont{Hutson}(2010)}]{hutson2010ultracold}
\bibinfo{author}{\bibfnamefont{J.~M.} \bibnamefont{Hutson}},
  \bibinfo{journal}{Science} \textbf{\bibinfo{volume}{327}},
  \bibinfo{pages}{788} (\bibinfo{year}{2010}).

\bibitem[{\citenamefont{Balakrishnan}(2016)}]{balakrishnan2016perspective}
\bibinfo{author}{\bibfnamefont{N.}~\bibnamefont{Balakrishnan}},
  \bibinfo{journal}{J. Chem. Phys.} \textbf{\bibinfo{volume}{145}},
  \bibinfo{pages}{150901} (\bibinfo{year}{2016}).

\bibitem[{\citenamefont{Ian}(2008)}]{wm2008low}
\bibinfo{author}{\bibfnamefont{W.~S.} \bibnamefont{Ian}},
  \emph{\bibinfo{title}{Low temperatures and cold molecules}}
  (\bibinfo{publisher}{World Scientific}, \bibinfo{year}{2008}).

\bibitem[{\citenamefont{Yan et~al.}(2013)\citenamefont{Yan, Moses, Gadway,
  Covey, Hazzard, Rey, Jin, and Ye}}]{yan2013observation}
\bibinfo{author}{\bibfnamefont{B.}~\bibnamefont{Yan}},
  \bibinfo{author}{\bibfnamefont{S.~A.} \bibnamefont{Moses}},
  \bibinfo{author}{\bibfnamefont{B.}~\bibnamefont{Gadway}},
  \bibinfo{author}{\bibfnamefont{J.~P.} \bibnamefont{Covey}},
  \bibinfo{author}{\bibfnamefont{K.~R.~A.} \bibnamefont{Hazzard}},
  \bibinfo{author}{\bibfnamefont{A.~M.} \bibnamefont{Rey}},
  \bibinfo{author}{\bibfnamefont{D.~S.} \bibnamefont{Jin}}, \bibnamefont{and}
  \bibinfo{author}{\bibfnamefont{J.}~\bibnamefont{Ye}},
  \bibinfo{journal}{Nature} \textbf{\bibinfo{volume}{501}},
  \bibinfo{pages}{521} (\bibinfo{year}{2013}).

\bibitem[{\citenamefont{Hazzard et~al.}(2014)\citenamefont{Hazzard, Gadway,
  Foss-Feig, Yan, Moses, Covey, Yao, Lukin, Ye, Jin et~al.}}]{hazzard2014many}
\bibinfo{author}{\bibfnamefont{K.~R.~A.} \bibnamefont{Hazzard}},
  \bibinfo{author}{\bibfnamefont{B.}~\bibnamefont{Gadway}},
  \bibinfo{author}{\bibfnamefont{M.}~\bibnamefont{Foss-Feig}},
  \bibinfo{author}{\bibfnamefont{B.}~\bibnamefont{Yan}},
  \bibinfo{author}{\bibfnamefont{S.~A.} \bibnamefont{Moses}},
  \bibinfo{author}{\bibfnamefont{J.~P.} \bibnamefont{Covey}},
  \bibinfo{author}{\bibfnamefont{N.~Y.} \bibnamefont{Yao}},
  \bibinfo{author}{\bibfnamefont{M.~D.} \bibnamefont{Lukin}},
  \bibinfo{author}{\bibfnamefont{J.}~\bibnamefont{Ye}},
  \bibinfo{author}{\bibfnamefont{D.~S.} \bibnamefont{Jin}},
  \bibnamefont{et~al.}, \bibinfo{journal}{Phys. Rev. Lett.}
  \textbf{\bibinfo{volume}{113}}, \bibinfo{pages}{195302}
  (\bibinfo{year}{2014}).

\bibitem[{\citenamefont{Schmidt and Lemeshko}(2015)}]{schmidt2015rotation}
\bibinfo{author}{\bibfnamefont{R.}~\bibnamefont{Schmidt}} \bibnamefont{and}
  \bibinfo{author}{\bibfnamefont{M.}~\bibnamefont{Lemeshko}},
  \bibinfo{journal}{Phys. Rev. Lett.} \textbf{\bibinfo{volume}{114}},
  \bibinfo{pages}{203001} (\bibinfo{year}{2015}).

\bibitem[{\citenamefont{Schmidt and Lemeshko}(2016)}]{schmidt2016deformation}
\bibinfo{author}{\bibfnamefont{R.}~\bibnamefont{Schmidt}} \bibnamefont{and}
  \bibinfo{author}{\bibfnamefont{M.}~\bibnamefont{Lemeshko}},
  \bibinfo{journal}{Phys. Rev. X} \textbf{\bibinfo{volume}{6}},
  \bibinfo{pages}{011012} (\bibinfo{year}{2016}).

\bibitem[{\citenamefont{Ni et~al.}(2018)\citenamefont{Ni, Rosenband, and
  Grimes}}]{ni2018}
\bibinfo{author}{\bibfnamefont{K.-K.} \bibnamefont{Ni}},
  \bibinfo{author}{\bibfnamefont{T.}~\bibnamefont{Rosenband}},
  \bibnamefont{and} \bibinfo{author}{\bibfnamefont{D.~D.}
  \bibnamefont{Grimes}}, \bibinfo{journal}{Chem. Sci.}
  \textbf{\bibinfo{volume}{9}}, \bibinfo{pages}{6830} (\bibinfo{year}{2018}).

\bibitem[{\citenamefont{Hudson and Campbell}(2018)}]{hudson2018dipolar}
\bibinfo{author}{\bibfnamefont{E.~R.} \bibnamefont{Hudson}} \bibnamefont{and}
  \bibinfo{author}{\bibfnamefont{W.~C.} \bibnamefont{Campbell}},
  \bibinfo{journal}{Phys. Rev. A} \textbf{\bibinfo{volume}{98}},
  \bibinfo{pages}{040302(R)} (\bibinfo{year}{2018}).

\bibitem[{\citenamefont{Campbell and Hudson}(2019)}]{campbell2019dipole}
\bibinfo{author}{\bibfnamefont{W.~C.} \bibnamefont{Campbell}} \bibnamefont{and}
  \bibinfo{author}{\bibfnamefont{E.~R.} \bibnamefont{Hudson}},
  \bibinfo{journal}{arXiv preprint arXiv:1909.02668}  (\bibinfo{year}{2019}).

\bibitem[{\citenamefont{Safronova et~al.}(2018)\citenamefont{Safronova, Budker,
  DeMille, Kimball, Derevianko, and Clark}}]{safronova2018search}
\bibinfo{author}{\bibfnamefont{M.~S.} \bibnamefont{Safronova}},
  \bibinfo{author}{\bibfnamefont{D.}~\bibnamefont{Budker}},
  \bibinfo{author}{\bibfnamefont{D.}~\bibnamefont{DeMille}},
  \bibinfo{author}{\bibfnamefont{D.~F.~J.} \bibnamefont{Kimball}},
  \bibinfo{author}{\bibfnamefont{A.}~\bibnamefont{Derevianko}},
  \bibnamefont{and} \bibinfo{author}{\bibfnamefont{C.~W.} \bibnamefont{Clark}},
  \bibinfo{journal}{Rev. Mod. Phys.} \textbf{\bibinfo{volume}{90}},
  \bibinfo{pages}{025008} (\bibinfo{year}{2018}).

\bibitem[{\citenamefont{Cairncross et~al.}(2017)\citenamefont{Cairncross,
  Gresh, Grau, Cossel, Roussy, Ni, Zhou, Ye, and Cornell}}]{Cairncross2017}
\bibinfo{author}{\bibfnamefont{W.~B.} \bibnamefont{Cairncross}},
  \bibinfo{author}{\bibfnamefont{D.~N.} \bibnamefont{Gresh}},
  \bibinfo{author}{\bibfnamefont{M.}~\bibnamefont{Grau}},
  \bibinfo{author}{\bibfnamefont{K.~C.} \bibnamefont{Cossel}},
  \bibinfo{author}{\bibfnamefont{T.~S.} \bibnamefont{Roussy}},
  \bibinfo{author}{\bibfnamefont{Y.}~\bibnamefont{Ni}},
  \bibinfo{author}{\bibfnamefont{Y.}~\bibnamefont{Zhou}},
  \bibinfo{author}{\bibfnamefont{J.}~\bibnamefont{Ye}}, \bibnamefont{and}
  \bibinfo{author}{\bibfnamefont{E.~A.} \bibnamefont{Cornell}},
  \bibinfo{journal}{Phys. Rev. Lett.} \textbf{\bibinfo{volume}{119}},
  \bibinfo{pages}{153001} (\bibinfo{year}{2017}).

\bibitem[{\citenamefont{Andreev et~al.}(2018)\citenamefont{Andreev, Ang,
  DeMille, Doyle, Gabrielse, Haefner, Hutzler, Lasner, Meisenhelder, O’Leary
  et~al.}}]{Andreev2018}
\bibinfo{author}{\bibfnamefont{V.}~\bibnamefont{Andreev}},
  \bibinfo{author}{\bibfnamefont{D.~G.} \bibnamefont{Ang}},
  \bibinfo{author}{\bibfnamefont{D.}~\bibnamefont{DeMille}},
  \bibinfo{author}{\bibfnamefont{J.~M.} \bibnamefont{Doyle}},
  \bibinfo{author}{\bibfnamefont{G.}~\bibnamefont{Gabrielse}},
  \bibinfo{author}{\bibfnamefont{J.}~\bibnamefont{Haefner}},
  \bibinfo{author}{\bibfnamefont{N.~R.} \bibnamefont{Hutzler}},
  \bibinfo{author}{\bibfnamefont{Z.}~\bibnamefont{Lasner}},
  \bibinfo{author}{\bibfnamefont{C.}~\bibnamefont{Meisenhelder}},
  \bibinfo{author}{\bibfnamefont{B.~R.} \bibnamefont{O’Leary}},
  \bibnamefont{et~al.}, \bibinfo{journal}{Nature}
  \textbf{\bibinfo{volume}{562}}, \bibinfo{pages}{355} (\bibinfo{year}{2018}).

\bibitem[{\citenamefont{Alighanbari et~al.}(2020)\citenamefont{Alighanbari,
  Giri, Constantin, Korobov, and Schiller}}]{Alighanbari2020}
\bibinfo{author}{\bibfnamefont{S.}~\bibnamefont{Alighanbari}},
  \bibinfo{author}{\bibfnamefont{G.~S.} \bibnamefont{Giri}},
  \bibinfo{author}{\bibfnamefont{F.~L.} \bibnamefont{Constantin}},
  \bibinfo{author}{\bibfnamefont{V.~I.} \bibnamefont{Korobov}},
  \bibnamefont{and} \bibinfo{author}{\bibfnamefont{S.}~\bibnamefont{Schiller}},
  \bibinfo{journal}{Nature} \textbf{\bibinfo{volume}{581}},
  \bibinfo{pages}{152} (\bibinfo{year}{2020}).

\bibitem[{\citenamefont{Chou et~al.}(2020)\citenamefont{Chou, Collopy, Kurz,
  Lin, Harding, Plessow, Fortier, Diddams, Leibfried, and
  Leibrandt}}]{Chou2020}
\bibinfo{author}{\bibfnamefont{C.~W.} \bibnamefont{Chou}},
  \bibinfo{author}{\bibfnamefont{A.~L.} \bibnamefont{Collopy}},
  \bibinfo{author}{\bibfnamefont{C.}~\bibnamefont{Kurz}},
  \bibinfo{author}{\bibfnamefont{Y.}~\bibnamefont{Lin}},
  \bibinfo{author}{\bibfnamefont{M.~E.} \bibnamefont{Harding}},
  \bibinfo{author}{\bibfnamefont{P.~N.} \bibnamefont{Plessow}},
  \bibinfo{author}{\bibfnamefont{T.}~\bibnamefont{Fortier}},
  \bibinfo{author}{\bibfnamefont{S.}~\bibnamefont{Diddams}},
  \bibinfo{author}{\bibfnamefont{D.}~\bibnamefont{Leibfried}},
  \bibnamefont{and} \bibinfo{author}{\bibfnamefont{D.~R.}
  \bibnamefont{Leibrandt}}, \bibinfo{journal}{Science}
  \textbf{\bibinfo{volume}{367}}, \bibinfo{pages}{1458} (\bibinfo{year}{2020}).

\bibitem[{\citenamefont{Viteau et~al.}(2008)\citenamefont{Viteau, Chotia,
  Allegrini, Bouloufa, Dulieu, Comparat, and Pillet}}]{viteau2008optical}
\bibinfo{author}{\bibfnamefont{M.}~\bibnamefont{Viteau}},
  \bibinfo{author}{\bibfnamefont{A.}~\bibnamefont{Chotia}},
  \bibinfo{author}{\bibfnamefont{M.}~\bibnamefont{Allegrini}},
  \bibinfo{author}{\bibfnamefont{N.}~\bibnamefont{Bouloufa}},
  \bibinfo{author}{\bibfnamefont{O.}~\bibnamefont{Dulieu}},
  \bibinfo{author}{\bibfnamefont{D.}~\bibnamefont{Comparat}}, \bibnamefont{and}
  \bibinfo{author}{\bibfnamefont{P.}~\bibnamefont{Pillet}},
  \bibinfo{journal}{Science} \textbf{\bibinfo{volume}{321}},
  \bibinfo{pages}{232} (\bibinfo{year}{2008}).

\bibitem[{\citenamefont{Staanum et~al.}(2010)\citenamefont{Staanum,
  H{\o}jbjerre, Skyt, Hansen, and Drewsen}}]{staanum2010rotational}
\bibinfo{author}{\bibfnamefont{P.~F.} \bibnamefont{Staanum}},
  \bibinfo{author}{\bibfnamefont{K.}~\bibnamefont{H{\o}jbjerre}},
  \bibinfo{author}{\bibfnamefont{P.~S.} \bibnamefont{Skyt}},
  \bibinfo{author}{\bibfnamefont{A.~K.} \bibnamefont{Hansen}},
  \bibnamefont{and} \bibinfo{author}{\bibfnamefont{M.}~\bibnamefont{Drewsen}},
  \bibinfo{journal}{Nat. Phys.} \textbf{\bibinfo{volume}{6}},
  \bibinfo{pages}{271} (\bibinfo{year}{2010}).

\bibitem[{\citenamefont{Schneider et~al.}(2010)\citenamefont{Schneider, Roth,
  Duncker, Ernsting, and Schiller}}]{schneider2010all}
\bibinfo{author}{\bibfnamefont{T.}~\bibnamefont{Schneider}},
  \bibinfo{author}{\bibfnamefont{B.}~\bibnamefont{Roth}},
  \bibinfo{author}{\bibfnamefont{H.}~\bibnamefont{Duncker}},
  \bibinfo{author}{\bibfnamefont{I.}~\bibnamefont{Ernsting}}, \bibnamefont{and}
  \bibinfo{author}{\bibfnamefont{S.}~\bibnamefont{Schiller}},
  \bibinfo{journal}{Nat. Phys.} \textbf{\bibinfo{volume}{6}},
  \bibinfo{pages}{275} (\bibinfo{year}{2010}).

\bibitem[{\citenamefont{Cournol et~al.}(2018)\citenamefont{Cournol, Pillet,
  Lignier, and Comparat}}]{cournol2018rovibrational}
\bibinfo{author}{\bibfnamefont{A.}~\bibnamefont{Cournol}},
  \bibinfo{author}{\bibfnamefont{P.}~\bibnamefont{Pillet}},
  \bibinfo{author}{\bibfnamefont{H.}~\bibnamefont{Lignier}}, \bibnamefont{and}
  \bibinfo{author}{\bibfnamefont{D.}~\bibnamefont{Comparat}},
  \bibinfo{journal}{Phys. Rev. A} \textbf{\bibinfo{volume}{97}},
  \bibinfo{pages}{031401(R)} (\bibinfo{year}{2018}).

\bibitem[{\citenamefont{Lien et~al.}(2014)\citenamefont{Lien, Seck, Lin,
  Nguyen, Tabor, and Odom}}]{lien2014broadband}
\bibinfo{author}{\bibfnamefont{C.-Y.} \bibnamefont{Lien}},
  \bibinfo{author}{\bibfnamefont{C.~M.} \bibnamefont{Seck}},
  \bibinfo{author}{\bibfnamefont{Y.-W.} \bibnamefont{Lin}},
  \bibinfo{author}{\bibfnamefont{J.~H.} \bibnamefont{Nguyen}},
  \bibinfo{author}{\bibfnamefont{D.~A.} \bibnamefont{Tabor}}, \bibnamefont{and}
  \bibinfo{author}{\bibfnamefont{B.~C.} \bibnamefont{Odom}},
  \bibinfo{journal}{Nat. Commun.} \textbf{\bibinfo{volume}{5}},
  \bibinfo{pages}{4783} (\bibinfo{year}{2014}).

\bibitem[{\citenamefont{Rellergert et~al.}(2013)\citenamefont{Rellergert,
  Sullivan, Schowalter, Kotochigova, Chen, and Hudson}}]{Rellergert2013}
\bibinfo{author}{\bibfnamefont{W.~G.} \bibnamefont{Rellergert}},
  \bibinfo{author}{\bibfnamefont{S.~T.} \bibnamefont{Sullivan}},
  \bibinfo{author}{\bibfnamefont{S.~J.} \bibnamefont{Schowalter}},
  \bibinfo{author}{\bibfnamefont{S.}~\bibnamefont{Kotochigova}},
  \bibinfo{author}{\bibfnamefont{K.}~\bibnamefont{Chen}}, \bibnamefont{and}
  \bibinfo{author}{\bibfnamefont{E.~R.} \bibnamefont{Hudson}},
  \bibinfo{journal}{Nature} \textbf{\bibinfo{volume}{495}},
  \bibinfo{pages}{490} (\bibinfo{year}{2013}), ISSN \bibinfo{issn}{1476-4687}.

\bibitem[{\citenamefont{Hansen et~al.}(2014)\citenamefont{Hansen, Versolato,
  Kristensen, Gingell, Schwarz, Windberger, Ullrich, L{\'o}pez-Urrutia, Drewsen
  et~al.}}]{hansen2014efficient}
\bibinfo{author}{\bibfnamefont{A.~K.} \bibnamefont{Hansen}},
  \bibinfo{author}{\bibfnamefont{O.}~\bibnamefont{Versolato}},
  \bibinfo{author}{\bibfnamefont{S.~B.} \bibnamefont{Kristensen}},
  \bibinfo{author}{\bibfnamefont{A.}~\bibnamefont{Gingell}},
  \bibinfo{author}{\bibfnamefont{M.}~\bibnamefont{Schwarz}},
  \bibinfo{author}{\bibfnamefont{A.}~\bibnamefont{Windberger}},
  \bibinfo{author}{\bibfnamefont{J.}~\bibnamefont{Ullrich}},
  \bibinfo{author}{\bibfnamefont{J.~C.} \bibnamefont{L{\'o}pez-Urrutia}},
  \bibinfo{author}{\bibfnamefont{M.}~\bibnamefont{Drewsen}},
  \bibnamefont{et~al.}, \bibinfo{journal}{Nature}
  \textbf{\bibinfo{volume}{508}}, \bibinfo{pages}{76} (\bibinfo{year}{2014}).

\bibitem[{\citenamefont{Tong et~al.}(2010)\citenamefont{Tong, Winney, and
  Willitsch}}]{tong2010sympathetic}
\bibinfo{author}{\bibfnamefont{X.}~\bibnamefont{Tong}},
  \bibinfo{author}{\bibfnamefont{A.~H.} \bibnamefont{Winney}},
  \bibnamefont{and}
  \bibinfo{author}{\bibfnamefont{S.}~\bibnamefont{Willitsch}},
  \bibinfo{journal}{Phys. Rev. Lett.} \textbf{\bibinfo{volume}{105}},
  \bibinfo{pages}{143001} (\bibinfo{year}{2010}).

\bibitem[{\citenamefont{Chou et~al.}(2017)\citenamefont{Chou, Kurz, Hume,
  Plessow, Leibrandt, and Leibfried}}]{chou2017preparation}
\bibinfo{author}{\bibfnamefont{C.-w.} \bibnamefont{Chou}},
  \bibinfo{author}{\bibfnamefont{C.}~\bibnamefont{Kurz}},
  \bibinfo{author}{\bibfnamefont{D.~B.} \bibnamefont{Hume}},
  \bibinfo{author}{\bibfnamefont{P.~N.} \bibnamefont{Plessow}},
  \bibinfo{author}{\bibfnamefont{D.~R.} \bibnamefont{Leibrandt}},
  \bibnamefont{and}
  \bibinfo{author}{\bibfnamefont{D.}~\bibnamefont{Leibfried}},
  \bibinfo{journal}{Nature} \textbf{\bibinfo{volume}{545}},
  \bibinfo{pages}{203} (\bibinfo{year}{2017}).

\bibitem[{\citenamefont{Shagam and Narevicius}(2013)}]{shagam2013sub}
\bibinfo{author}{\bibfnamefont{Y.}~\bibnamefont{Shagam}} \bibnamefont{and}
  \bibinfo{author}{\bibfnamefont{E.}~\bibnamefont{Narevicius}},
  \bibinfo{journal}{The Journal of Physical Chemistry C}
  \textbf{\bibinfo{volume}{117}}, \bibinfo{pages}{22454}
  (\bibinfo{year}{2013}).

\bibitem[{\citenamefont{Ni et~al.}(2008)\citenamefont{Ni, Ospelkaus,
  De~Miranda, Pe'Er, Neyenhuis, Zirbel, Kotochigova, Julienne, Jin, and
  Ye}}]{ni2008high}
\bibinfo{author}{\bibfnamefont{K.-K.} \bibnamefont{Ni}},
  \bibinfo{author}{\bibfnamefont{S.}~\bibnamefont{Ospelkaus}},
  \bibinfo{author}{\bibfnamefont{M.}~\bibnamefont{De~Miranda}},
  \bibinfo{author}{\bibfnamefont{A.}~\bibnamefont{Pe'Er}},
  \bibinfo{author}{\bibfnamefont{B.}~\bibnamefont{Neyenhuis}},
  \bibinfo{author}{\bibfnamefont{J.}~\bibnamefont{Zirbel}},
  \bibinfo{author}{\bibfnamefont{S.}~\bibnamefont{Kotochigova}},
  \bibinfo{author}{\bibfnamefont{P.}~\bibnamefont{Julienne}},
  \bibinfo{author}{\bibfnamefont{D.}~\bibnamefont{Jin}}, \bibnamefont{and}
  \bibinfo{author}{\bibfnamefont{J.}~\bibnamefont{Ye}},
  \bibinfo{journal}{Science} \textbf{\bibinfo{volume}{322}},
  \bibinfo{pages}{231} (\bibinfo{year}{2008}).

\bibitem[{\citenamefont{Bressel et~al.}(2012)\citenamefont{Bressel, Borodin,
  Shen, Hansen, Ernsting, and Schiller}}]{bressel2012manipulation}
\bibinfo{author}{\bibfnamefont{U.}~\bibnamefont{Bressel}},
  \bibinfo{author}{\bibfnamefont{A.}~\bibnamefont{Borodin}},
  \bibinfo{author}{\bibfnamefont{J.}~\bibnamefont{Shen}},
  \bibinfo{author}{\bibfnamefont{M.}~\bibnamefont{Hansen}},
  \bibinfo{author}{\bibfnamefont{I.}~\bibnamefont{Ernsting}}, \bibnamefont{and}
  \bibinfo{author}{\bibfnamefont{S.}~\bibnamefont{Schiller}},
  \bibinfo{journal}{Phys. Rev. Lett.} \textbf{\bibinfo{volume}{108}},
  \bibinfo{pages}{183003} (\bibinfo{year}{2012}).

\bibitem[{\citenamefont{Werner et~al.}(1982)\citenamefont{Werner, Rosmus, and
  Grimm}}]{werner_ab_1982}
\bibinfo{author}{\bibfnamefont{H.-J.} \bibnamefont{Werner}},
  \bibinfo{author}{\bibfnamefont{P.}~\bibnamefont{Rosmus}}, \bibnamefont{and}
  \bibinfo{author}{\bibfnamefont{M.}~\bibnamefont{Grimm}},
  \bibinfo{journal}{Chem. Phys.} \textbf{\bibinfo{volume}{73}},
  \bibinfo{pages}{169} (\bibinfo{year}{1982}).

\bibitem[{\citenamefont{Cai and Fran{\c{c}}ois}(1999)}]{cai1999ab}
\bibinfo{author}{\bibfnamefont{Z.-L.} \bibnamefont{Cai}} \bibnamefont{and}
  \bibinfo{author}{\bibfnamefont{J.-P.} \bibnamefont{Fran{\c{c}}ois}},
  \bibinfo{journal}{J. Mol. Spectrosc.} \textbf{\bibinfo{volume}{197}},
  \bibinfo{pages}{12} (\bibinfo{year}{1999}).

\bibitem[{\citenamefont{Chattopadhyaya
  et~al.}(2003)\citenamefont{Chattopadhyaya, Chattopadhyay, and
  Das}}]{chattopadhyaya2003electronic}
\bibinfo{author}{\bibfnamefont{S.}~\bibnamefont{Chattopadhyaya}},
  \bibinfo{author}{\bibfnamefont{A.}~\bibnamefont{Chattopadhyay}},
  \bibnamefont{and} \bibinfo{author}{\bibfnamefont{K.~K.} \bibnamefont{Das}},
  \bibinfo{journal}{J. Mol. Struct. THEOCHEM} \textbf{\bibinfo{volume}{639}},
  \bibinfo{pages}{177} (\bibinfo{year}{2003}).

\bibitem[{\citenamefont{Li et~al.}(2019)\citenamefont{Li, Yuan, Liang, Wu,
  Wang, and Yan}}]{li2019laser}
\bibinfo{author}{\bibfnamefont{R.}~\bibnamefont{Li}},
  \bibinfo{author}{\bibfnamefont{X.}~\bibnamefont{Yuan}},
  \bibinfo{author}{\bibfnamefont{G.}~\bibnamefont{Liang}},
  \bibinfo{author}{\bibfnamefont{Y.}~\bibnamefont{Wu}},
  \bibinfo{author}{\bibfnamefont{J.}~\bibnamefont{Wang}}, \bibnamefont{and}
  \bibinfo{author}{\bibfnamefont{B.}~\bibnamefont{Yan}},
  \bibinfo{journal}{Chem. Phys.} \textbf{\bibinfo{volume}{525}},
  \bibinfo{pages}{110412} (\bibinfo{year}{2019}).

\bibitem[{\citenamefont{Di~Rosa}(2004)}]{DiRosa2004}
\bibinfo{author}{\bibfnamefont{M.~D.} \bibnamefont{Di~Rosa}},
  \bibinfo{journal}{Eur. Phys. J. D} \textbf{\bibinfo{volume}{31}},
  \bibinfo{pages}{395} (\bibinfo{year}{2004}), ISSN \bibinfo{issn}{1434-6079}.

\bibitem[{\citenamefont{Sofikitis et~al.}(2009)\citenamefont{Sofikitis, Weber,
  Fioretti, Horchani, Allegrini, Chatel, Comparat, and
  Pillet}}]{sofikitis2009molecular}
\bibinfo{author}{\bibfnamefont{D.}~\bibnamefont{Sofikitis}},
  \bibinfo{author}{\bibfnamefont{S.}~\bibnamefont{Weber}},
  \bibinfo{author}{\bibfnamefont{A.}~\bibnamefont{Fioretti}},
  \bibinfo{author}{\bibfnamefont{R.}~\bibnamefont{Horchani}},
  \bibinfo{author}{\bibfnamefont{M.}~\bibnamefont{Allegrini}},
  \bibinfo{author}{\bibfnamefont{B.}~\bibnamefont{Chatel}},
  \bibinfo{author}{\bibfnamefont{D.}~\bibnamefont{Comparat}}, \bibnamefont{and}
  \bibinfo{author}{\bibfnamefont{P.}~\bibnamefont{Pillet}},
  \bibinfo{journal}{New J. Phys.} \textbf{\bibinfo{volume}{11}},
  \bibinfo{pages}{055037} (\bibinfo{year}{2009}).

\bibitem[{\citenamefont{Nguyen and Odom}(2011)}]{nguyen2011prospects}
\bibinfo{author}{\bibfnamefont{J.~H.~V.} \bibnamefont{Nguyen}}
  \bibnamefont{and} \bibinfo{author}{\bibfnamefont{B.}~\bibnamefont{Odom}},
  \bibinfo{journal}{Phys. Rev. A} \textbf{\bibinfo{volume}{83}},
  \bibinfo{pages}{053404} (\bibinfo{year}{2011}).

\bibitem[{\citenamefont{Nguyen et~al.}(2011)\citenamefont{Nguyen, Viteri,
  Hohenstein, Sherrill, Brown, and Odom}}]{nguyen2011challenges}
\bibinfo{author}{\bibfnamefont{J.~H.~V.} \bibnamefont{Nguyen}},
  \bibinfo{author}{\bibfnamefont{C.~R.} \bibnamefont{Viteri}},
  \bibinfo{author}{\bibfnamefont{E.~G.} \bibnamefont{Hohenstein}},
  \bibinfo{author}{\bibfnamefont{C.~D.} \bibnamefont{Sherrill}},
  \bibinfo{author}{\bibfnamefont{K.~R.} \bibnamefont{Brown}}, \bibnamefont{and}
  \bibinfo{author}{\bibfnamefont{B.}~\bibnamefont{Odom}}, \bibinfo{journal}{New
  J. Phys.} \textbf{\bibinfo{volume}{13}}, \bibinfo{pages}{063023}
  (\bibinfo{year}{2011}).

\bibitem[{\citenamefont{Stollenwerk et~al.}(2017)\citenamefont{Stollenwerk,
  Odom, Kokkin, and Steimle}}]{stollenwerk2017electronic}
\bibinfo{author}{\bibfnamefont{P.~R.} \bibnamefont{Stollenwerk}},
  \bibinfo{author}{\bibfnamefont{B.~C.} \bibnamefont{Odom}},
  \bibinfo{author}{\bibfnamefont{D.~L.} \bibnamefont{Kokkin}},
  \bibnamefont{and} \bibinfo{author}{\bibfnamefont{T.}~\bibnamefont{Steimle}},
  \bibinfo{journal}{J. Mol. Spectrosc.} \textbf{\bibinfo{volume}{332}},
  \bibinfo{pages}{26} (\bibinfo{year}{2017}).

\bibitem[{\citenamefont{Lin and Odom}(2016)}]{lin2016high}
\bibinfo{author}{\bibfnamefont{Y.-W.} \bibnamefont{Lin}} \bibnamefont{and}
  \bibinfo{author}{\bibfnamefont{B.~C.} \bibnamefont{Odom}},
  \bibinfo{journal}{arXiv:1610.04324}  (\bibinfo{year}{2016}).

\bibitem[{\citenamefont{Lien et~al.}(2011)\citenamefont{Lien, Williams, and
  Odom}}]{lien2011optical}
\bibinfo{author}{\bibfnamefont{C.-Y.} \bibnamefont{Lien}},
  \bibinfo{author}{\bibfnamefont{S.~R.} \bibnamefont{Williams}},
  \bibnamefont{and} \bibinfo{author}{\bibfnamefont{B.}~\bibnamefont{Odom}},
  \bibinfo{journal}{Phys. Chem. Chem. Phys.} \textbf{\bibinfo{volume}{13}},
  \bibinfo{pages}{18825} (\bibinfo{year}{2011}).

\bibitem[{\citenamefont{Stollenwerk et~al.}(2019)\citenamefont{Stollenwerk,
  Antonov, and Odom}}]{stollenwerk2017ip}
\bibinfo{author}{\bibfnamefont{P.~R.} \bibnamefont{Stollenwerk}},
  \bibinfo{author}{\bibfnamefont{I.~O.} \bibnamefont{Antonov}},
  \bibnamefont{and} \bibinfo{author}{\bibfnamefont{B.~C.} \bibnamefont{Odom}},
  \bibinfo{journal}{J. Mol. Spectrosc.} \textbf{\bibinfo{volume}{355}},
  \bibinfo{pages}{40} (\bibinfo{year}{2019}).

\bibitem[{\citenamefont{Baba and Waki}(2001)}]{baba2001laser}
\bibinfo{author}{\bibfnamefont{T.}~\bibnamefont{Baba}} \bibnamefont{and}
  \bibinfo{author}{\bibfnamefont{I.}~\bibnamefont{Waki}}, \bibinfo{journal}{J.
  Appl. Phys.} \textbf{\bibinfo{volume}{89}}, \bibinfo{pages}{4592}
  (\bibinfo{year}{2001}).

\bibitem[{\citenamefont{Honjou}(2003)}]{honjou2003ab}
\bibinfo{author}{\bibfnamefont{N.}~\bibnamefont{Honjou}},
  \bibinfo{journal}{Mol. Phys.} \textbf{\bibinfo{volume}{101}},
  \bibinfo{pages}{3063} (\bibinfo{year}{2003}).

\bibitem[{\citenamefont{Shi et~al.}(2012)\citenamefont{Shi, Li, Xing, Sun, Zhu,
  and Liu}}]{shi2012mrci}
\bibinfo{author}{\bibfnamefont{D.}~\bibnamefont{Shi}},
  \bibinfo{author}{\bibfnamefont{W.}~\bibnamefont{Li}},
  \bibinfo{author}{\bibfnamefont{W.}~\bibnamefont{Xing}},
  \bibinfo{author}{\bibfnamefont{J.}~\bibnamefont{Sun}},
  \bibinfo{author}{\bibfnamefont{Z.}~\bibnamefont{Zhu}}, \bibnamefont{and}
  \bibinfo{author}{\bibfnamefont{Y.}~\bibnamefont{Liu}},
  \bibinfo{journal}{Comput. Theor. Chem.} \textbf{\bibinfo{volume}{980}},
  \bibinfo{pages}{73} (\bibinfo{year}{2012}).

\bibitem[{Sci(manuscript in preparation)}]{Science2020}
 (\bibinfo{year}{manuscript in preparation}).

\bibitem[{\citenamefont{Western}(2017)}]{western2017pgopher}
\bibinfo{author}{\bibfnamefont{C.~M.} \bibnamefont{Western}},
  \bibinfo{journal}{J. Quant. Spectrosc. Radiat. Transfer}
  \textbf{\bibinfo{volume}{186}}, \bibinfo{pages}{221} (\bibinfo{year}{2017}).

\bibitem[{\citenamefont{Willits et~al.}(2012)\citenamefont{Willits, Weiner, and
  Cundiff}}]{willits2012line}
\bibinfo{author}{\bibfnamefont{J.~T.} \bibnamefont{Willits}},
  \bibinfo{author}{\bibfnamefont{A.~M.} \bibnamefont{Weiner}},
  \bibnamefont{and} \bibinfo{author}{\bibfnamefont{S.~T.}
  \bibnamefont{Cundiff}}, \bibinfo{journal}{Opt. Express}
  \textbf{\bibinfo{volume}{20}}, \bibinfo{pages}{3110} (\bibinfo{year}{2012}).

\bibitem[{\citenamefont{Scholl et~al.}(1995)\citenamefont{Scholl, Cameron,
  Rosner, and Holt}}]{scholl_laser-rf_1995}
\bibinfo{author}{\bibfnamefont{T.~J.} \bibnamefont{Scholl}},
  \bibinfo{author}{\bibfnamefont{R.}~\bibnamefont{Cameron}},
  \bibinfo{author}{\bibfnamefont{S.~D.} \bibnamefont{Rosner}},
  \bibnamefont{and} \bibinfo{author}{\bibfnamefont{R.~A.} \bibnamefont{Holt}},
  \bibinfo{journal}{Can. J. Phys.} \textbf{\bibinfo{volume}{73}},
  \bibinfo{pages}{101} (\bibinfo{year}{1995}).

\bibitem[{\citenamefont{Stollenwerk et~al.}(2018)\citenamefont{Stollenwerk,
  Kokish, de~Oliveira-Filho, Ornellas, and Odom}}]{stollenwerk2018optical}
\bibinfo{author}{\bibfnamefont{P.~R.} \bibnamefont{Stollenwerk}},
  \bibinfo{author}{\bibfnamefont{M.~G.} \bibnamefont{Kokish}},
  \bibinfo{author}{\bibfnamefont{A.~G.~S.} \bibnamefont{de~Oliveira-Filho}},
  \bibinfo{author}{\bibfnamefont{F.~R.} \bibnamefont{Ornellas}},
  \bibnamefont{and} \bibinfo{author}{\bibfnamefont{B.~C.} \bibnamefont{Odom}},
  \bibinfo{journal}{Atoms} \textbf{\bibinfo{volume}{6}}, \bibinfo{pages}{53}
  (\bibinfo{year}{2018}).

\end{thebibliography}

\end{document}


\title{Supplementary Material}

\maketitle

\section{What Happens to Population Pumped to $v>0$?}
Our experimental observation that there is not significant population trapped in $v>0$ can be understood from the SiO\+ spectrum. Off-diagonal decays during optical pumping will result in some population being pumped to $v>0$.  The set of transitions $\ket{X, v} \rightarrow \ket{B, v}$ are denoted $(v,v)$ bands.  Since the vibrational frequency ($\omega_e$) of the B state (1138~cm$^{-1}$) is slightly less than that of the X state (1162~cm$^{-1}$), the $(v,v)$ band origins are red-detuned by $\sim$25 cm$^{-1}$ for each subsequent $v$. Consequently, the broadband spectrum we use for cooling to $N=0$ covers the P-branch as well as the low-$N$ portion of the R-branch for the first few $(v,v)$ bands with $v>0$~\cite{Science2020}.

Once a molecule is pumped to $v=1$, the optical pumping light will continue to drive transitions in the $(1,1)$ band.  The $v=1$ lifetime has not yet been measured but has been calculated to be $>1$~s~\cite{nguyen2011prospects}, so off-diagonal decay (most often to $v=0$ or $v=2$) is by far the dominant exit channel from $(1,1)$ cycling.

The situation for $(2,2)$ cycling is different. The energy of $\ket{X, v=2}$ is only 90 cm$^{-1}$ above $\ket{A, v=0}$. Due to this near-degeneracy and resultant experimentally verified strong coupling\,\cite{rosner_study_1998}, $\ket{B, v=2} \rightarrow \ket{A, v=0}$ borrows $\sim$10\% of the (2,2) band intensity.  This results in a significant enhancement (we calculate $\Gamma \sim 1\times10^{6}\ \textrm{s}^{-1}$) over other $\ket{B} \rightarrow \ket{A}$ decay channels ($\Gamma \sim 2\times10^{3}\ \textrm{s}^{-1}$~\cite{li2019laser}).  Calculations predict that population in \ket{A,v=0} subsequently decays to \ket{X,v=0} in 7 ms~\cite{li2019laser}.

Since for \ket{X,v>2} decay to $A$ is also energetically allowed, the calculated transition moments predict that they quickly decay to $\ket{A,v}$ states and from there quickly return to \ket{X,v<2}, with the timescale for both decays being $<10$~ms. We therefore do not expect significant population in \ket{X,v>2}, because the rate of pumping to these states is much slower than the rate of their decay.

\section{$N=0$ Cooling Model}
For population beginning in an even parity state, optical pumping down the rotational ladder can be modeled by a rate $\lambda$ for population transfer from $N$ to $N-2$, where $\lambda$ varies by less than a factor of 2 across the $P$-branch~\cite{stollenwerk-thesis}. Neglecting this weak dependence, we arrive at a simple analytic expression for pumping of population initially in $N = 2M$ into $N=0$:
\begin{equation}
\label{coolfit}
n_0(t) = 1 - e^{-\lambda t}\sum_{i=0}^{M-1} \frac{(\lambda t)^i}{i!}.\\
\end{equation}

Experimentally, we observe that population beginning in odd parity states eventually flips parity and is cooled to $N=0$, but the mechanism is not yet understood.  Magnetic dipole transitions are too slow to explain the observed parity-flip timescale. The process likely starts from population cooled to the lowest negative parity state ($N=1$) being continuously driven on the quasi-cycling $P(1)$ transition, until it leaks elsewhere. The required odd number of electric-dipole transitions either occur in decay channels passing through intermediate \ket{X, v>0} or \ket{A} states. Off-diagonal decay from $\ket{B, v=0} \rightarrow \ket{X, v>0}$ occurs faster than the observed parity-flip timescale, but subsequent \ket{X, v>0} vibrational relaxation is much slower than the observed timescale.  Likewise, although the predicted lifetime of \ket{A} is itself short enough, decay from $\ket{B, v=0} \rightarrow \ket{A}$ is predicted to have too small a branching ratio~\cite{li2019laser}.   However, a hybrid mechanism seems plausible, shown in Fig.~1(c) of the main text.  In this sequence, population in \ket{X, v=2} continues to be driven to \ket{B,v=2} by the cooling laser and eventually undergoes a decay $\ket{B, v=2} \rightarrow \ket{A, v=0}$, as discussed above.  Relaxation from $\ket{A, v=0} \rightarrow \ket{X}$ is expected to occur on the order of 7 ms~\cite{li2019laser}.

Regardless of the parity flip mechanism, by modeling the process as governed by an effective rate $\lambda_P$, population transfer from $N=2M+1$ to $N=0$ is analytically found to be
\begin{equation}
\label{coolfitodd}
n_0(t) = 1 - e^{-\lambda t}\left(\sum_{i=0}^M \frac{(\lambda t)^i}{i!}+\sum_{i=M+1}^{\infty} \frac{(\lambda - \lambda_P)^it^i}{i!(1-\lambda_P/\lambda)^M}\right).\\
\end{equation}
For a slow parity flip timescale $\lambda_P\ll \lambda/M$, the result asymptotically approaches $1-e^{-\lambda_Pt}$. In the limit $\lambda_P \rightarrow \lambda$, the second sum goes to zero and we recover Eq.\,\eqref{coolfit} with an additional step on the ladder.

\begin{figure}
\includegraphics[width=3.375in]{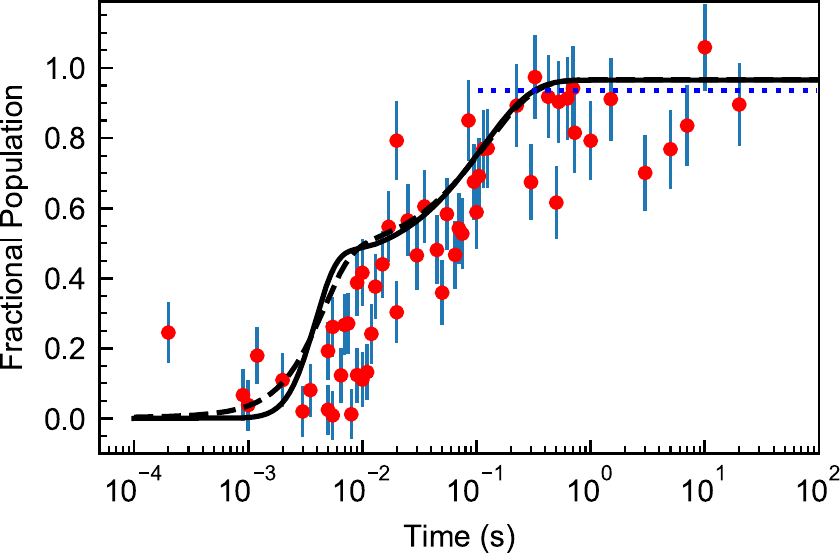}
\caption{\label{fig:sm_cool} Simulation of the cooling to $N=0$ (dashed line) and fit of the model to the simulation (solid line).  Also shown are the same data from Fig.~4 of the main text and the asymptote of their fit after correction for systematic shifts.}
\end{figure}

The cooling data are fit to a sum of Eqns.\,\eqref{coolfit} and \eqref{coolfitodd}, with the results shown in Fig.\,4 of the main text and Table\,\ref{tab:temp_syst}. We also used an Einstein rate equation simulation~\cite{stollenwerk2018optical} to simulate the cooling (Fig.\,\ref{fig:sm_cool}). The simulation uses the values of the experimental spectral intensity and spectral cutoff sharpness of the cooling laser, with SiO\+\ initialized to a 300 K distribution.  The simulation curve was also fit using the model.

The cooling timescale predicted by the simulation is noticeably shorter than the data for the even-parity cooling step.  The simulation models the cooling laser spectrum as a continuum. The simulation models the cooling laser spectrum as continuum. In reality, the spectrum of the femtosecond laser is a frequency comb with 80 MHz spacing between the frequency peaks. The frequency offset of the peaks is uncontrolled. The detuning of cooling transitions from comb teeth can be as high as 40 MHz, significantly larger than the 2 MHz natural linewidth of the $B-X$ transition. However Doppler broadening from thermal motion and from micromotion mitigates the detuning by some degree, which we have not yet characterized. (To fully correct this problem of spectral gaps in the future, the laser could be modulated to fill in the gaps in spectral intensity.)  The uncontrolled comb structure of the cooling laser could explain the timescale discrepancy between the simulation and the data, but further characterization of the tooth stability and Doppler-broadened SiO\+ spectrum would be required to test this hypothesis.

\section{Systematic Shifts in Cooling Asymptote}
There are known systematic shifts which affect the measured value of the $N=0$ cooling asymptote. SiO$^+$ reacts with background H$_2$ \,\cite{fahey1981reactions}, so the maximum possible dissociation fraction of SiO$^+$ is reduced from 1. The most abundant isotopologue is $^{28}$Si$^{16}$O with a natural abundance of 92$\%$. The shifts of $^{29}$Si$^{16}$O\+ and $^{30}$Si$^{16}$O\+ are expected to be 0.51 cm$^{-1}$ and 0.99 cm$^{-1}$ respectively; thus, probability of their dissociation is reduced. Finally, $^{28}$Si$^{16}$O$^+$ transitions originating from $N=1$ are roughly one linewidth away and also contribute to the dissociated fraction. The details of these systematics are discussed in~\cite{stollenwerk-thesis}.
Accounting for the systematic effects, we conclude that the steady-state $N=0$ absolute population fraction of the dominant isotope is 0.94(3), which corresponds to a temperature of 0.53(6) K. This result is in agreement with both the simulation and the relative population analysis.

\begin{table}[]
\caption{Absolute cooling efficiency fitted parameters, systematic shifts, and results.\\}
\label{tab:temp_syst}
\begin{tabular}{lccccc}
\hline
\hline
\multicolumn{3}{c|}{\textbf{Cooling Rate Parameters}} &
  \multicolumn{3}{c}{\textbf{Systematic Effects}} \\ \hline
\textbf{}              & \textbf{Data} & \multicolumn{1}{c|}{\textbf{Sim.}}            & \textbf{}          & \textbf{$\Delta A$} & \textbf{Unc.} \\ \hline
$\lambda$ (s$^{-1}$)   & 650(90)       & \multicolumn{1}{c|}{1630} & Isotope            & -0.07               & $<0.002$      \\
                       &               & \multicolumn{1}{c|}{}     & Reaction           & -0.038              & 0.010         \\
$\lambda_P$ (s$^{-1}$) & 9(4)          & \multicolumn{1}{c|}{8.9}  & $n_1$              & +0.006              & $<0.005$      \\
                       &               & \multicolumn{1}{c|}{}     & Total $\Delta A$ & $\overline{-0.102}$              & $\overline{0.011}$         \\ \hline
\multicolumn{6}{c}{\textbf{Absolute Cooling Efficiency}}                                                                                                    \\
$n_0\footnotemark[1]$                      & 0.94(3) &                      &$n_{0,sim}$             &0.97 &  \\
$T$ (K)                                    & 0.53(6) &                   &  $T _{sim} $(K)            &0.47  &  \\
\hline \hline
\end{tabular}
\footnotetext[1]{Asymptotic $N=0$ population using $n_0$ = $A - \Delta A$, where $A$ is the asymptotic value from the fits to data.}
\end{table}

\section{Broadband Pumping of Non-Diagonal Molecules}
It is interesting to consider optical pumping on arbitrary molecules. Use of VIPA could directly extend the technique discussed here to diagonal-FCF molecules of heavier reduced mass. The development of broadband and spectrally bright supercontinuum lasers may eventually allow efficient optical pumping of non-diagonal molecules with a very small number of repump lasers. Furthermore, replacing the spectral mask with a digital
micro-mirror array~\cite{horchani2015femtosecond} will allow for construction of more complex masks which can vary in time, potentially guiding population through spectrally congested regions with conflicting selection rules~\cite{Science2020}.
